\documentclass[aps,prb,reprint,superscriptaddress]{revtex4-1}

\usepackage{graphicx}

\begin{document}

\title{Determination of the magnetic structure of CePt$_2$In$_7$ by means of neutron diffraction}

\author{M. Raba}
\email[]{matthias.raba@lncmi.cnrs.fr}
\affiliation{Laboratoire National des Champs Magn\'{e}tiques Intenses (LNCMI-EMFL), CNRS, UGA, F-38042 Grenoble, France}
\affiliation{Institut N\'{e}el, Universit\'{e} Grenoble Alpes, F-38000 Grenoble, France}
\affiliation{CNRS, Institut N\'{e}el, F-38000 Grenoble, France}

\author{E. Ressouche}
\affiliation{INAC, CEA and Universit\'{e} Grenoble Alpes, CEA Grenoble, F-38054 Grenoble, France}

\author{N. Qureshi}
\affiliation{Institut Laue Langevin, 71 rue des Martyrs, BP156, F-38042 Grenoble Cedex 9, France}

\author{C. V. Colin}
\affiliation{Institut N\'{e}el, Universit\'{e} Grenoble Alpes, F-38000 Grenoble, France}
\affiliation{CNRS, Institut N\'{e}el, F-38000 Grenoble, France}

\author{V. Nassif}
\affiliation{Institut N\'{e}el, Universit\'{e} Grenoble Alpes, F-38000 Grenoble, France}
\affiliation{CNRS, Institut N\'{e}el, F-38000 Grenoble, France}

\author{S. Ota}
\affiliation{Graduate School of Science and Technology, Niigata University, Niigata 950-2181, Japan}

\author{Y. Hirose}
\affiliation{Department of Physics, Niigata University, Niigata 950-2181, Japan}

\author{R. Settai}
\affiliation{Department of Physics, Niigata University, Niigata 950-2181, Japan}

\author{P. Rodi\`{e}re}
\email[]{pierre.rodiere@neel.cnrs.fr}
\affiliation{Institut N\'{e}el, Universit\'{e} Grenoble Alpes, F-38000 Grenoble, France}
\affiliation{CNRS, Institut N\'{e}el, F-38000 Grenoble, France}

\author{I. Sheikin}
\email[]{ilya.sheikin@lncmi.cnrs.fr}
\affiliation{Laboratoire National des Champs Magn\'{e}tiques Intenses (LNCMI-EMFL), CNRS, UGA, F-38042 Grenoble, France}

\date{\today}

\begin{abstract}
The magnetic structure of the heavy fermion antiferromagnet CePt$_2$In$_7$ is determined using neutron diffraction. We find a magnetic wave vector $\mathbf{q}_M = (1/2,1/2,1/2)$, which is temperature independent up to $T_N =$ 5.5~K. A staggered moment of 0.45(1)$\mu_B$ at 2~K resides on the Ce ion. The nearest-neighbor moments in the tetragonal basal plane are aligned antiferromagnetically. The moments rotate by 90$^\circ$ from one CeIn$_3$ plane to another along the $c$ axis. A much weaker satellite peak with an incommensurate magnetic wave vector $\mathbf{q}_M = (1/2,1/2,0.47)$ seems to develop at low temperature. However, the experimental data available so far are not sufficient to draw a definitive conclusion about the possible co-existence of commensurate and incommensurate magnetic structures in this material.
\end{abstract}

\maketitle


CePt$_2$In$_7$ is a recently discovered heavy fermion compound that belongs to the same family as the well-studied CeIn$_3$ and Ce$M$In$_5$ ($M =$ Co, Rh, Ir). The spacing between Ce-In planes in CePt$_2$In$_7$ is drastically increased~\cite{Klimczuk2014} as compared to its Ce$M$In$_5$ counterparts, implying a more two-dimensional crystal structure. Expectedly, the Fermi surface of CePt$_2$In$_7$ is also much more two dimensional~\cite{Altarawneh2011}. This compound crystallizes in the body-centered-tetragonal structure (space group $I4/mmm$) with a unit cell considerably elongated along the $c$ axis.

CePt$_2$In$_7$ undergoes an antiferromagnetic (AF) transition at $T_N =$ 5.5~K~\cite{Tobash2012,Bauer2010a,Warren2010,Warren2012}. Recent electrical resistivity and ac-calorimetry measurements under pressure on single crystals of CePt$_2$In$_7$ revealed a quantum critical point at a critical pressure $P_c \approx$ 3.2~GPa, where the AF order is completely suppressed~\cite{Sidorov2013}. A superconducting dome with the highest transition temperature $T_c \simeq$ 2.1~K is observed around $P_c$~\cite{Sidorov2013,Bauer2010,Bauer2010a}, suggesting that critical AF fluctuations may mediate the Cooper pairing. Nuclear magnetic and quadrupole resonance (NMR and NQR, respectively) measurements on single crystals under pressure~\cite{Sakai2014} suggest that $P_c$ is not the only relevant pressure for this material. Indeed, a localized to itinerant crossover of the 4$f$ electron of Ce occurs within the AF state at $P^* \approx$ 2.4~GPa, approximately the pressure where superconductivity first emerges in single crystals.

While the magnetic structure of the cubic CeIn$_3$ is characterized by a simple commensurate ordering wave vector $(1/2, 1/2, 1/2)$~\cite{Lawrence1980,Benoit1980}, that of the more two-dimensional CeRhIn$_5$ is more complicated. Its magnetically ordered ground state is an incommensurate helicoidal phase with the propagation vector $\mathbf{q}_M = (1/2,1/2,0.297)$ and the magnetic moment in the basal plane of the tetragonal structure~\cite{Bao2000,Bao2003,Raymond2007,Raymond2014,Fobes2017}.

In CePt$_2$In$_7$, the magnetic structure of its AF ground state is still an open question. The existing reports on this matter are controversial. Indeed, NQR studies performed on polycrystalline samples~\cite{Warren2010} suggest that antiferromagnetism is commensurate in this material. The same conclusion was drawn from positive muon-spin rotation and relaxation ($\mu^+$SR) measurements also performed on polycrystalline samples~\cite{Mansson2014}. On the contrary, the NQR spectra obtained on single crystals are consistent with the coexistence of an incommensurate and commensurate AF component of the magnetic structure~\cite{Sakai2011,Sakai2014}. The commensurate AF order first occurs just below $T_N$, then, at lower temperatures of about 3 K, incommensurate AF order gradually grows in. This probably accounts for the observation of only commensurate magnetism in polycrystalline NQR measurements, which were performed at temperatures down to 4~K~\cite{Warren2010}. At 1.6~K, the volume fraction of the incommensurate order is about 75\%. However, the commensurate AF order is stabilized by hydrostatic pressure: Its volume fraction becomes nearly 100\% at 2.4~GPa, the pressure where superconductivity first occurs and $f$ electrons change from localized to itinerant. All the NQR experiments~\cite{Warren2010,Sakai2011,Sakai2014} lead to the same conclusion: The magnetic propagation vector is $(1/2,1/2,\delta)$, although the value of $\delta$ is not predicted.

\begin{figure}[htb]
\includegraphics[width=8cm]{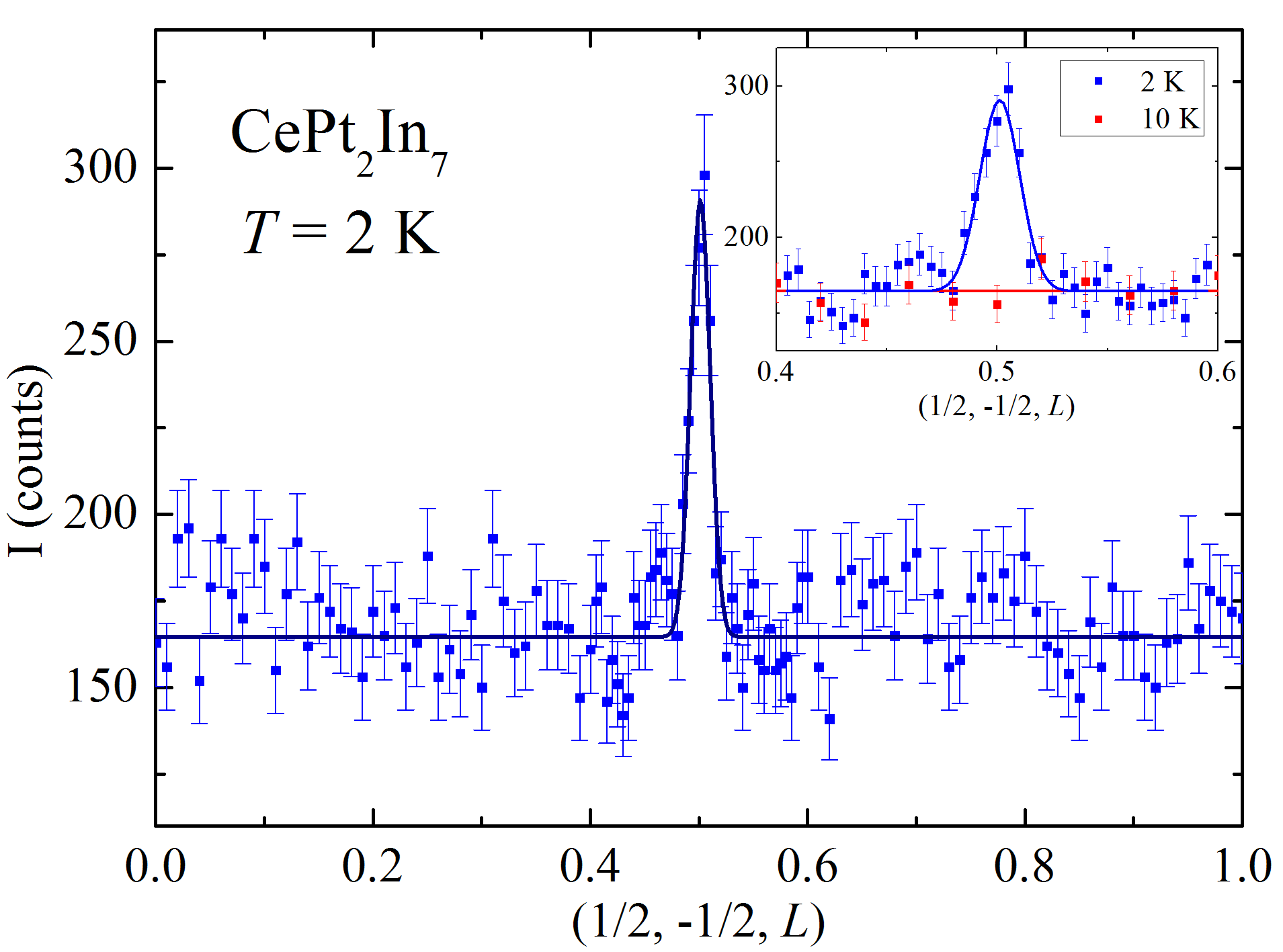}
\caption{\label{Q_scan}$\mathbf{Q}$ scan performed along the [0,0,1] direction at 2 K. The inset shows a zoom of the same $\mathbf{Q}$ scan near $\mathbf{Q} = (1/2, -1/2, 1/2)$ at 2 and 10~K. The intensity is in number of counts per $4\times10^6$ monitor counts, which corresponds roughly to 7 min. The solid lines are Gaussian fits of the peaks.}
\end{figure}

In this Rapid Communication, we report the magnetic structure of CePt$_2$In$_7$ determined from neutron diffraction, which is a bulk probe contrary to NQR and $\mu^+$SR measurements. The magnetic structure is found to be commensurate with a magnetic propagation vector (1/2, 1/2, 1/2). The magnetic moments are aligned antiferromagnetically in the basal plane, and rotate by 90$^\circ$ from one CeIn$_3$ plane to another. As expected, the magnetic structure is more two dimensional than those of either CeIn$_3$ or CeRhIn$_5$.


Single crystals of CePt$_2$In$_7$ were grown by the In self-flux method, as explained in more detail elsewhere~\cite{Kurahashi2015}. The high quality of the samples is confirmed by specific heat measurements that show a clear and unique second-order phase transition at the N\'{e}el temperature $T_N =$ 5.5~K~\cite{Krupko2016} and de Haas-van Alphen effect measurements exhibiting quantum oscillations starting from about 2~T~\cite{Goetze}. Two neutron diffraction experiments were performed at the Institut Laue-Langevin (ILL) of Grenoble (France) to determine the AF structure of CePt$_2$In$_7$. At first, a powder neutron diffraction was carried out on the high-flux diffractometer D1B by using a pyrolytic graphite (002) monochromator providing a beam with a wavelength of 2.52~\AA{ }and a 128$^\circ$ multidetector. A great number of single crystals was ground into 1.6~g of fine powder. These measurements did not reveal any magnetic peaks below T$_N$ in spite of 10-h-long acquisitions at both 1.5 and 10~K. This puts an upper limit for the magnetic moment at about 0.8$\mu_B$. Both the crystal structure and the lattice parameters remain unchanged in the AF state. Similarly, no temperature-dependent magnetic peaks were detected in the previous powder neutron diffraction measurements on CeRhIn$_5$~\cite{Bao2000}. Following this, a preliminary single-crystal neutron diffraction experiment was performed on the D23 beamline. This test measurement revealed a magnetic peak with $\mathbf{q}_M = (1/2,1/2,1/2)$, leading to more detailed measurements on the D10 beamline. For this experiment, we examined the biggest available sample from the same batch with the dimensions $1.9 \times 1.9 \times 1.0$~mm$^3$ and the $c$ axis perpendicular to the platelet surface. The instrument was used in a four-circle configuration with an additional triple-axis energy analysis. The latter was used in the elastic mode with a single detector to increase the signal-to-noise ratio. A vertically focusing pyrolytic graphite monochromator and analyzer was employed, fixing the incident and analyzed wavelength to 2.36~\AA. A pyrolytic graphite filter was used to reduce the higher harmonic contamination to $10^{-4}$ of the primary beam intensity. In order to reach temperatures down to 2~K, we used a four-circle cryostat with helium circulation. The measured lattice parameters at $T =$ 2~K are $a =$ 4.595(2)~\AA{ }and $c =$ 21.558(5)~\AA, as obtained from the analysis of 30 nuclear Bragg peaks.


\begin{figure}[htb]
\includegraphics[width=8cm]{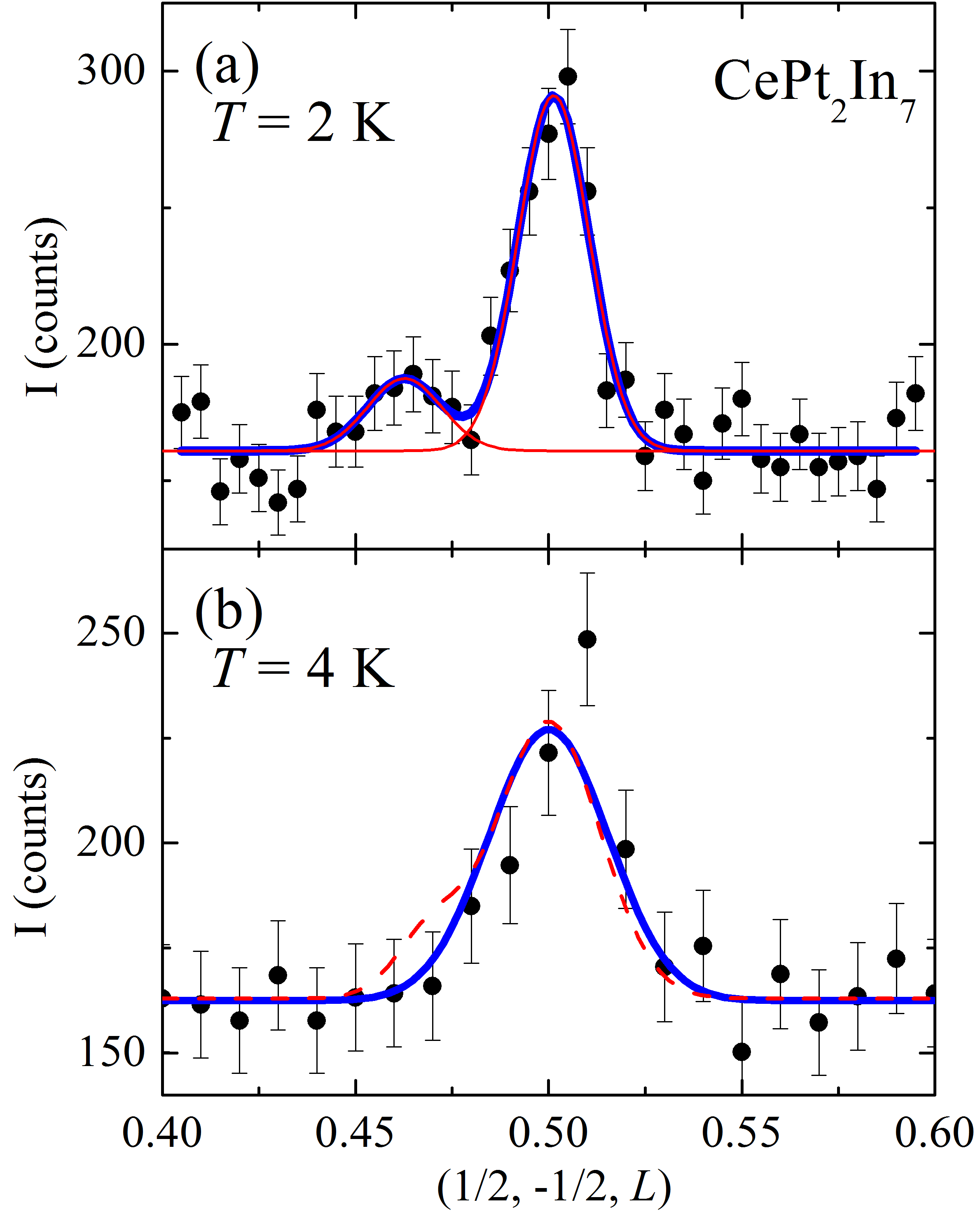}
\caption{\label{Double_peak}Detailed $\mathbf{Q}$ scans performed along the [0,0,1] direction at (a) 2~K and (b) 4~K. The intensity is in number of counts per $4\times10^6$ monitor counts, which corresponds roughly to 7 min. The lines are Gaussian fits of the peaks. The data obtained at 2~K are well fit by two Gaussian peaks centered at (1/2, -1/2, 1/2) and (1/2, -1/2, 0.47). At 4~K, an attempt to fit the data by two Gaussian peaks while fixing the position of the second peak and the integrated intensity ratio of the two peaks (dashed line) does not yield a satisfactory result. The data are much better fitted by a single Gaussian peak (solid line).}
\end{figure}

Figure~\ref{Q_scan} shows the $\mathbf{Q}$ scan performed along the [001] direction at $T =$ 2~K. A clear peak is observed at $\mathbf{Q} = (1/2,1/2,1/2)$. As expected, the $(1/2,1/2,1/2)$ Bragg peak disappears above the N\'{e}el temperature, as shown in the inset of Fig.~\ref{Q_scan} for $T =$ 10~K. Below $T_N$, the $q_l$ value of the propagation vector does not seem to change with temperature. However, the full width at half maximum (FWHM) of the magnetic peak increases with temperature (see Fig.~\ref{Double_peak}), suggesting a smaller size of magnetic domains along the $c$ axis for temperatures close to $T_N$. We did not observe any other obvious peaks with an incommensurate magnetic wave vector. However, a small satellite peak at (1/2, 1/2, 1/2-$\delta$) with $\delta$=0.03 seems to emerge above the background noise at $T =$ 2~K, as shown in Fig.~\ref{Double_peak}(a). The integrated intensity of this peak is 21\% of the commensurate magnetic Bragg peak. The satellite peak seems to disappear at $T =$ 4~K, as an attempt to fit the data by two Gaussian peaks with the fixed position of the incommensurate peak and integrated intensity ratio of the two peaks does not yield a satisfactory result [see Fig.~\ref{Double_peak}(b)]. Remarkably, the NQR measurements performed on single crystals suggest the presence of only a commensurate order at $T =$ 4~K, whereas an incommensurate order starts to develop below about 3~K~\cite{Sakai2011,Sakai2014}. However, these measurements also suggest that the volume fraction of the incommensurate order is almost 3/4 at 2~K, and that the internal field due to the incommensurate order is larger than that due to its commensurate counterpart. This is difficult to reconcile with our results, if the small satellite peak was a signature of an additional incommensurate order. Furthermore, if this was the case, a peak at (1/2,1/2,1/2+$\delta$) should have also been observed due to the tetragonal symmetry, although not necessarily with the same intensity. Such a peak, however, is absent in our data. Therefore, for the rest of this Rapid Communication, we will neglect this tiny contribution, whose possible magnetic origin remains to be clarified.

\begin{figure}[htb]
\includegraphics[width=8cm]{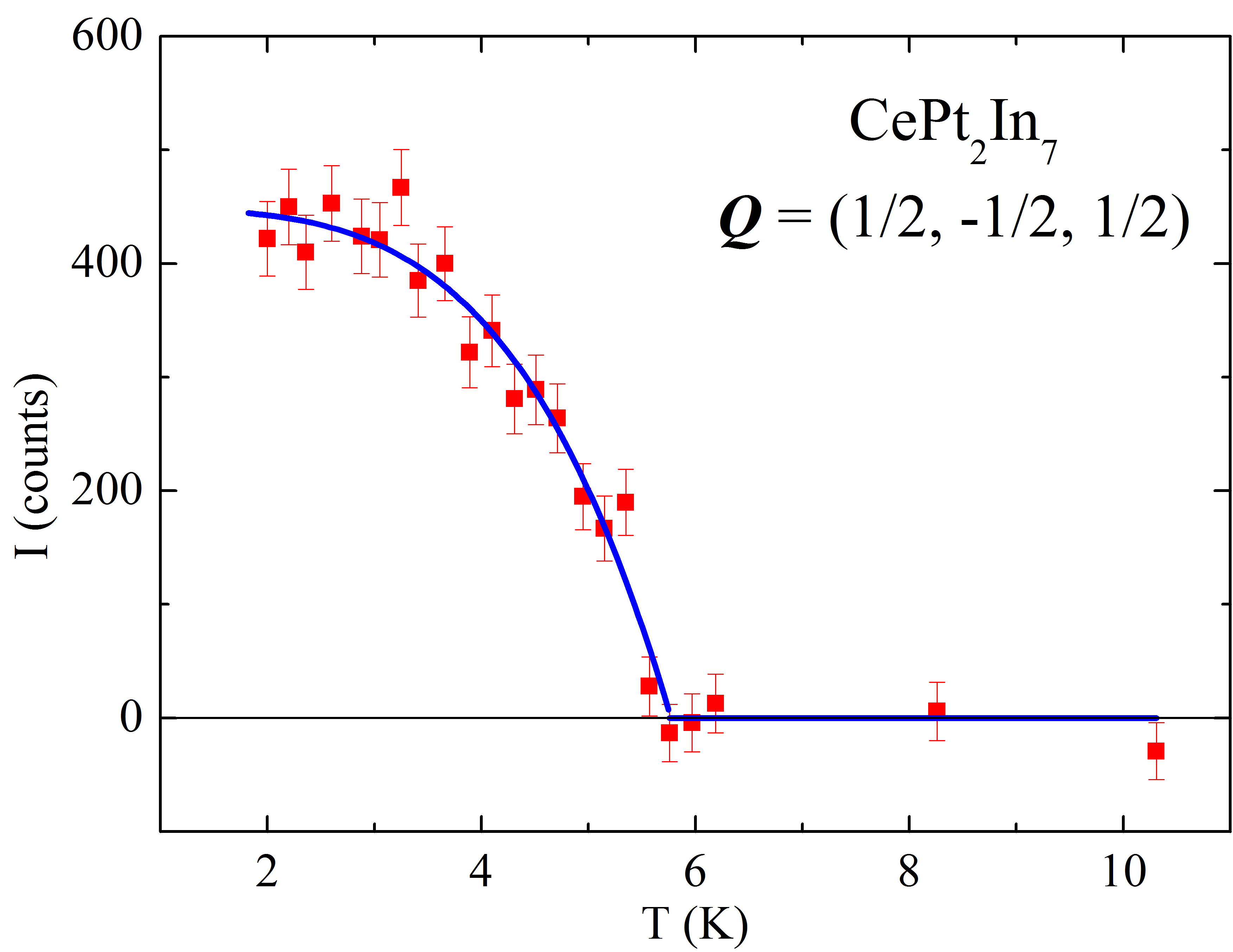}
\caption{\label{Temperature_dependence}Temperature dependence of the (1/2, -1/2, 1/2) magnetic Bragg peak intensity after subtracting the background. The intensity is in number of counts per $1.6\times10^7$ monitor counts, which corresponds roughly to 27 min. The line is a phenomenological fit as explained in the text.}
\end{figure}

Figure~\ref{Temperature_dependence} shows the temperature dependence of the $(1/2, -1/2, 1/2)$ magnetic Bragg peak intensity $I$ proportional to the square of the magnetic phase transition order parameter. To determine the N\'{e}el temperature, the data were fitted by a phenomenological function $I/I_0 = 1 - (T/T_N)^\alpha$, with $\alpha$ a free parameter, which does not have a particular physical meaning. This function was successfully used to fit the temperature dependence of the magnetic Bragg peak intensity in other heavy fermion compounds, such as CePd$_2$Si$_2$~\cite{Dijk2000,Kernavanois2005} and Sn-doped CeRhIn$_5$~\cite{Raymond2014}. The best fit is obtained with $\alpha = 4.1\pm0.7$ and $T_N = 5.7\pm0.1$~K. The latter value is consistent with $T_N$ determined from previous measurements~\cite{Tobash2012,Bauer2010a,Warren2010,Warren2012,Krupko2016}.

\begin{table}[hbt]
\caption{\label{tab:mag_peaks}Magnetic refinement for the magnetic structure discussed in the text and shown in Fig.~\ref{Magnetic_structure}. (Note: $\chi^2$ = 1.84).}
\begin{ruledtabular}
\begin{tabular}{c c c}
  $\mathbf{Q}$ & $I_{\mathrm{obs}}$ & $I_{\mathrm{calc}}$ \\
  \hline
  (1/2, -1/2, -1/2) & 8(1) & 8.2 \\
  (-1/2, -1/2, -1/2) & 7(1) & 8.2 \\
  (-1/2, -1/2, 1/2) & 7(1) & 8.2 \\
  (1/2, -1/2, 1/2) & 7.0(9) & 8.2 \\
  (1/2, -1/2, -5/2) & 10(1) & 10.6 \\
  (-1/2, -1/2, -5/2) & 9(1) & 10.6 \\
  (-1/2, -1/2, 5/2) & 11(1) & 10.6 \\
  (1/2, -1/2, 5/2) & 9(1) & 10.6 \\
  (1/2, -1/2, -3/2) & 10(1) & 9.3 \\
  (-1/2, -1/2, -3/2) & 9(1) & 9.3 \\
  (-1/2, -1/2, 3/2) & 8(1) & 9.3 \\
  (1/2, -1/2, 3/2) & 7(1) & 9.3 \\
  (1/2, -1/2, -9/2) & 12(2) & 12.0 \\
  (-1/2, -1/2, -9/2) & 19(3) & 12.0 \\
  (-1/2, -1/2, 9/2) & 13(2) & 12.0 \\
  (1/2, -1/2, 9/2) & 15(2) & 12.0 \\
  (1/2, -1/2, -7/2) & 14(2) & 11.5 \\
  (-1/2, -1/2, -7/2) & 13(2) & 11.5 \\
  (-1/2, -1/2, 7/2) & 12(1) & 11.5 \\
  (1/2, -1/2, 7/2) & 11(2) & 11.5 \\
  (1/2, -1/2, -11/2) & 15(2) & 12.0 \\
  (-1/2, -1/2, -11/2) & 18.8(25) & 12.0 \\
  (1/2, 1/2, 11/2) & 13(2) & 12.0 \\
  (-1/2, 1/2, 11/2) & 18(3) & 12.0 \\
\end{tabular}
\end{ruledtabular}
\end{table}

\begin{figure}[htb]
\vspace{5mm}
\includegraphics[width=\linewidth]{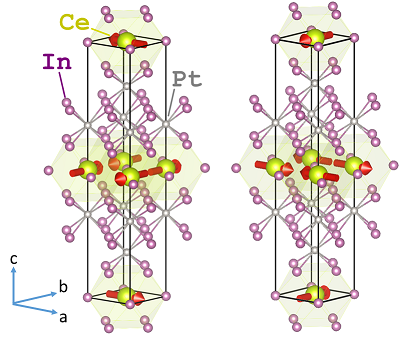}
\caption{\label{Magnetic_structure}Magnetic structure of CePt$_2$In$_7$ in a structural unit cell. The magnetic moment, schematically shown by arrows, is 0.45$\mu_B$ per Ce and it is aligned in the basal plane. Both magnetic domains are shown.}
\end{figure}

In order to determine the magnetic structure and the staggered moment, we measured 24 magnetic peaks at 2~K. For each peak, the measured neutron Bragg intensity was corrected for extinction, absorption and Lorentz factor; the resulting intensities are shown in Table~\ref{tab:mag_peaks}. Only two arrangements of the magnetic moments are allowed by the group theory analysis: They can be either aligned in the basal plane or along the $c$ axis.
The best refinement, taking into account two magnetic domains and assuming they are equally populated, is obtained for magnetic moments aligned antiferromagnetically in the basal plane. The moments rotate by 90$^\circ$ from one CeIn$_3$ plane to its nearest-neighbor along the $c$ axis. For this structure, the comparison between the observed intensities and the calculated ones is shown in Table~\ref{tab:mag_peaks}. Given the tetragonal symmetry of CePt$_2$In$_7$, we cannot determine the orientation of the magnetic moments in the basal plane. However, previous NQR data~\cite{Warren2010,Sakai2011} strongly suggest that they are parallel to the [100] direction in the commensurate phase. The resulting magnetic structure (magnetic space group $I_c\bar{4}2m$) is schematically shown in Fig.~\ref{Magnetic_structure}. The staggered magnetic moment is determined at 2~K to be $M = 0.45(1)\mu_B$ per Ce.

\begin{table}
\caption{\label{tab:parameters}Crystallographic and magnetic properties of Ce$T_m$In$_{3+2m}$ ($T$ = Rh, Pt) compounds. Here, $d$ is the distance between CeIn$_3$ planes, $\alpha$ is the angle between magnetic moments on neighboring CeIn$_3$ planes, and $M$ is the staggered moment.}
\begin{ruledtabular}
\begin{tabular}{c c c c c}
   & $m$ & $d$~(\AA) & $\alpha$ & $M$ ($\mu_B$/Ce) \\
  \hline
  CeIn$_3$ & 0 & 4.6 & 180$^\circ$ & 0.48~\cite{Benoit1980}, 0.65~\cite{Lawrence1980} \\
  CeRhIn$_5$ & 1 & 7.5 & 107$^\circ$ & 0.54~\cite{Fobes2017}, 0.59~\cite{Raymond2007,Raymond2014}, 0.75~\cite{Bao2000,Bao2003} \\
  CePt$_2$In$_7$ & 2 & 10.8 & 90$^\circ$ & 0.45 \\
\end{tabular}
\end{ruledtabular}
\end{table}


Having determined the magnetic structure of CePt$_2$In$_7$, we now consider the relationship between the magnetic and crystal structures within the Ce$T_m$In$_{3+2m}$ ($T =$ Rh, Pt) family, where $m$ $T$In$_2$ layers separate a single CeIn$_3$ layer (see Table~\ref{tab:parameters}). The building block of the family, CeIn$_3$ ($m = 0$), crystallizes into a simple cubic structure (space group $Pm\bar{3}m$). For CeRhIn$_5$ ($m = 1$) with alternating layers of CeIn$_3$ and RhIn$_2$, the crystal structure is primitive tetragonal (space group $P4/mmm$). Finally, CePt$_2$In$_7$ ($m = 2$), where two PtIn$_2$ layers separate each CeIn$_3$ layer, crystallizes into a body-centered-tetragonal structure with $I4/mmm$ space group. In all three materials, the magnetic moments of the Ce ions form a square lattice, surrounded by In ions in the $a-b$ plane. They all are simple, nearest-neighbor antiferromagnets in the plane. In CeIn$_3$, the magnetic moments are also arranged antiferromagnetically along the $c$ axis. In CeRhIn$_5$, magnetic correlations across the RhIn$_2$ layer are incommensurate, with neighboring magnetic moments being rotated by approximately 107$^\circ$~\cite{Bao2000,Bao2003}. In CePt$_2$In$_7$, the magnetic moments are also aligned in the $a-b$ plane. In this case, however, they rotate by 90$^\circ$ from one plane to another, as shown in Fig.~\ref{Magnetic_structure}. Clearly, the rotation angle is reduced with increasing the number of $T$In$_2$ layers between the CeIn$_3$ layers (see Table~\ref{tab:parameters}). The magnetic structure of CePt$_2$In$_7$ is expected to be the most two dimensional among the three compounds. Indeed, a greater separation between the CeIn$_3$ planes leads to a weaker coupling between them. In addition, geometrical frustration due to the body-centered-tetragonal crystal structure is likely to further reduce the effective dimensionality, as was experimentally observed in BaCuSi$_2$O$_6$~\cite{Sebastian2006}. The 90$^\circ$ rotation angle of the moments from one plane to another observed in CePt$_2$In$_7$ is probably a signature of its most two-dimensional magnetic structure.

The magnetic structure of both CeIn$_3$ and CePt$_2$In$_7$ is commensurate. The same conclusion was drawn for Ce$_2$RhIn$_8$, in which two layers of CeIn$_3$ are separated by a single layer of RhIn$_2$~\cite{Bao2001}. This suggests CeRhIn$_5$ with its incommensurate magnetic structure as a unique member of the whole Ce$_nT_m$In$_{3n+2m}$ ($T =$ transition metal) family, where $m$ $T$In$_2$ layers separate $n$ CeIn$_3$ layers. In all these compounds, superconductivity emerges in the vicinity of a quantum critical point induced either by pressure~\cite{Mathur1998,Nicklas2003,Knebel2006,Sidorov2013} or chemical doping~\cite{Pagliuso2001,Ohira-Kawamura2007,Yokoyama2008}. Interestingly, a commensurate magnetic order was observed to either co-exist or compete with incommensurate ordering in CeRhIn$_5$ doped with either Ir~\cite{Llobet2005} or Co~\cite{Ohira-Kawamura2007,Yokoyama2008}. Remarkably, in these compounds, commensurate antiferromagnetism emerges in the vicinity of a quantum critical point where superconductivity also appears. Furthermore, in Sn-doped CeRhIn$_5$, a drastic change in the magnetic order and a commensurate antiferromagnetism was observed in the proximity of the quantum critical point~\cite{Raymond2014} where superconductivity is expected, but has not been observed so far. This suggests that a commensurate magnetic order might be favorable for the formation of superconductivity around a quantum critical point in this family of materials. On the other hand, several neutron diffraction experiments performed in CeRhIn$_5$ under pressure up to 1.7~GPa did not reveal the presence of a commensurate AF order~\cite{Majumdar2002,Llobet2004,Raymond2008}. This pressure, however, is considerably lower than the critical value, $P_c \approx 2.4$~GPa, although a pressure-induced bulk superconductivity is observed above about 1.5 GPa~\cite{Knebel2006}.

Regarding the staggered moment, its values are comparable for all three compounds of the Ce$T_m$In$_{3+2m}$ family (see Table~\ref{tab:parameters}). Given that the staggered moment of Ce$_2$RhIn$_8$, $M =$ 0.55$\mu_B$/Ce~\cite{Bao2001}, is also of the same order, there is no obvious correlation between the staggered moment and dimensionality in this family of heavy fermion materials.

Finally, the magnetic propagation vector (1/2, 1/2, 1/2) observed both in CeIn$_3$ and CePt$_2$In$_7$ implies that the magnetic Brillouin zone in these compounds is eight times smaller than the crystallographic one. In CeIn$_3$, this naturally accounts for the observation of only small Fermi surfaces in quantum oscillation measurements performed at moderate magnetic fields up to 17 T~\cite{Endo2005}. Very high magnetic fields of about 55~T are required for the observation of large Fermi surfaces through magnetic breakdown tunneling~\cite{Harrison2007}. Similarly, only small Fermi surface pockets, well within the interior of the AF Brillouin zone, are observed in CePt$_2$In$_7$ at moderately high magnetic fields up to about 25-30~T~\cite{Miyake2015,Goetze}. This is in contrast with CeRhIn$_5$, where large Fermi surfaces are observed at relatively low fields below 17-18~T~\cite{Hall2001a,Shishido2002}.


In conclusion, we find the commensurate magnetic structure with $\mathbf{q}_M = (1/2, 1/2, 1/2)$ as depicted in Fig.~\ref{Magnetic_structure} for CePt$_2$In$_7$. A magnetic moment of 0.45(1)$\mu_B$ at 2~K resides on the Ce ion  and the basal plane is its easy plane. Within the basal plane, magnetic moments form a simple nearest-neighbor antiferromagnet on a square lattice. The moments rotate by 90$^\circ$ from one CeIn$_3$ plane to another along the $c$ axis.

\begin{acknowledgments}
We thank Y. Tokunaga, M. Horvatic, and J. Robert for fruitful discussions. We acknowledge scientific and technical support we received during the powder neutron diffraction experiment on the CRG-D1B (ILL) operated by the CNRS. This work was partially supported by the ANR-DFG grant ``Fermi-NESt''.
\end{acknowledgments}

\bibliography{CePt2In7}

\end{document}